\documentstyle[mprocl]{article}

\def\be{\begin{equation}}
\def\ee{\end{equation}}
\def\bea{\begin{eqnarray}}
\def\eea{\end{eqnarray}}

\begin{document}
\title{GRAVITATIONAL VACUUM POLARIZATION}
\author{MATT VISSER}
\address{Physics Department, Washington University, \\
Saint Louis, Missouri 63130--4899, USA}

\maketitle
\abstracts{
The energy conditions of classical Einstein gravity fail once quantum
effects are introduced.  These quantum violations of the energy
conditions are not subtle high-energy Planck scale effects. Rather the
quantum violations of the energy conditions already occur in
semiclassical quantum gravity and are first-order $O(\hbar)$
effects. Quantum violations of the energy conditions are widespread,
albeit small.
}
  
\section{Introduction}

The energy conditions of classical Einstein gravity are used to
prove many powerful and general theorems such as the singularity
theorems, the positive mass theorem, and the topological censorship
theorem. While quantum mechanical violation of the energy conditions
was certainly expected, it is perhaps a little surprising just how
widespread the quantum violations of the energy conditions are.
Quantum induced violations of the energy conditions are now known
to occur in:
\begin{itemize}
\item
Casimir effect.~\cite{Morris-Thorne,Visser}
\item
Squeezed vacuum.~\cite{Morris-Thorne,Visser}
\item
Cosmological inflation.~\cite{Visser}$^{\!-\,}$\cite{Zed15}
\item
Cosmological particle production.~\cite{Zeldovich}
\item 
Conformal anomaly.~\cite{anec}
\item
Gravitational vacuum polarization.~\cite{gvp1}$^{\!-\,}$\cite{gvp4}
\end{itemize}

\section{Gravitational vacuum polarization}

When a gravitational field acts as an external source applied to
a quantum field theory (QFT) it will distort the quantum vacuum,
and shift the expectation value of the stress-energy so that $\langle
T^{\mu\nu} \rangle \neq 0$.  Typically $\langle T^{\mu\nu} \rangle
\approx \hbar c/L^4$, where $L$ is some position-dependent characteristic
length scale, which may be associated either with the quantum state,
or with the background geometry.  If the gravitational field of
interest contains an event horizon then there are at least four
different natural definitions of the quantum mechanical vacuum
state:
\begin{itemize}
\item 
Hartle--Hawking vacuum [thermal equilibrium at infinity].
\item
Boulware vacuum [empty at infinity].
\item
Unruh vacuum [evaporating black hole].
\item
Vacuum cleaner vacuum [accreting black hole].
\end{itemize}
These different states correspond to different definitions of normal
ordering on the spacetime. If the spacetime does not possess an
event horizon (star, planet) then these particular complications
go away and you only have one vacuum state to deal with---the
Boulware vacuum. (Corresponding to normal ordering with respect to
the usual static $t$ coordinate.)

In contrast to QFT defined on flat Minkowski space, in curved space
no analytic calculations are currently possible.  One resorts either
to numerical methods~\cite{AHS,JLO91} or to nonperturbative analytic
approximations.  To keep matters tractable, I confine attention to
massless conformally coupled scalar QFT on the Schwarzschild
geometry. All calculations are performed in the test-field limit,
and back reaction is not included.  (For recent progress regarding
back reaction see Hochberg--Popov--Sushkov.~\cite{HPS})

\section{Violations in the Hartle--Hawking vacuum}

In the Hartle--Hawking vacuum, the energy condition violations are
confined to the region between the event horizon and the unstable
photon orbit at $r=3M$. The dominant energy condition (DEC) is the
first energy condition to be violated at $r=2.992 M$, next the weak
energy condition (WEC) is violated at $r=2.438M$, finally the null
energy condition (NEC) and strong energy condition (SEC) are violated
once $r=2.298M$. You can show this by using the numerically calculated
values of the stress-energy tensor.~\cite{gvp1,AHS} For consistency,
you can check that the same qualitative features survive when using
Page's analytic approximation.~\cite{gvp1} The fact that the DEC
violations first occur so close to the unstable photon orbit is
suggestive, but may merely be a numerical accident.

\section{Violations in the Boulware vacuum}

In the Boulware vacuum all the pointwise energy conditions are
violated throughout the entire region exterior to the event
horizon.~\cite{gvp2} This is proved by using the numerically
calculated values of the stress-energy tensor.~\cite{gvp2,AHS} Again,
for consistency, you can check that the same qualitative features
survive when using the Page--Brown--Ottewill analytic
approximation.~\cite{gvp2} This property should continue to hold
for the external region outside a star or planet, the presence of
the event horizon not being the crucial feature in deriving this
result.~\cite{gvp2} Working perturbatively around flat space,
Flanagan and Wald~\cite{Flanagan-Wald}  have verified that these
energy condition violations are generic to first order in $GM/r$.

\section{Violations in the Unruh vacuum}

In the Unruh vacuum all the pointwise energy conditions are violated
throughout the entire region exterior to the event horizon.~\cite{gvp4}
Again, this is best shown by using the numerically calculated values
of the stress-energy tensor.~\cite{gvp4}$^{\!-\,}$\cite{JLO91} You
have to be careful since testing the energy conditions involves
subtracting numbers that are numerically close to each other. At
large distances the stress-energy tensor is dominated by the Hawking
flux, but the Hawking flux quietly cancels out of the NEC evaluated
on outgoing null geodesics, so it is the subdominant pieces of the
stress-energy tensor that are critical in testing for energy
condition violations.  For good measure, no good analytic approximation
is known for the Unruh vacuum, and the numerical data are all that
one has to work with.~\cite{gvp4,JLO91}

\section{Violations in $(1+1)$ dimensions}

In $(1+1)$ dimensions the expectation value of the stress--energy can
be calculated analytically. The exact results obtained for this case
are qualitatively in agreement with the pattern found (numerically
and/or via analytic approximation) in $(3+1)$ dimensions.~\cite{gvp3}
This serves as a useful sanity check on the entire formalism.

\section{Discussion}

Quantum violations of the energy conditions are widespread, albeit
small. There are important issues of principle at stake: Present
versions of the positive mass theorem and the singularity theorems are
strictly classical and do not survive the introduction of even
semiclassical quantum effects.  (The hypotheses of these theorems are
known to be false in semiclassical quantum gravity.)  There are
suspicions that it might be possible to place general bounds on the
size of the quantum violations~\cite{Inequalities} and so still be
able to prove some type of generalized singularity theorem and/or
generalized positive mass theorem, but it is far from clear how to go
about this.

\section*{Acknowledgments}

This research was supported by the U.S. Department of Energy.

\section*{References}

\end{document}